\newcommand{\beq} {\begin{eqnarray}}
\newcommand{\eeq} {\end{eqnarray}}
\documentstyle[prd,aps,amstex,latexsym,epsfig,amsfonts]{revtex}
\def\edcomment#1{\iffalse\marginpar{\raggedright\sl#1\/}\else\relax\fi}
\reversemarginpar
\begin{document}

\title{ Difficulties with Re-collapsing Models in Closed Isotropic Loop Quantum Cosmology}

\author{Daniel Green and  William G. Unruh}
                         
\address{ 
Department of Physics and Astronomy\\
University of British Columbia\\
Vancouver, BC, Canada, V6T 1Z1
\\email: {\tt drgreen@@stanford.edu, unruh@@physics.ubc.ca}}

\maketitle

\begin{abstract}
The use of techniques from loop quantum gravity for cosmological models may solve some
difficult problems in quantum cosmology.  The solutions under a number of circumstances have
been well studied.  We will analyse the behaviour of solutions in the closed model, focusing on the
behaviour of a universe containing a massless scalar field.  The asymptotic behaviour of the solutions is
examined, and is used to determine requirements of the initial conditions.
\end{abstract}

\section{Introduction}

On large scales, the description of the universe provided by general relativity is consistent with all
measurements to date.  The cosmological models that have been generated seem to account for
much of the history of the universe.  However, the cosmological singularity that appears in this
approach has suggested that a quantum theory of cosmology is required.  Most studies of
quantum cosmology have involved mini-superspace models.  Homogeneity (and
sometimes isotropy) is used to limit the degrees of freedom of the classical framework to a finite
number.  The reduced phase space is then quantized.  Unfortunately, most such quantizations have been
unable to provide a consistent solution to the singularity and initial value problems.  Furthermore,
the absence of time has lead to additional interpretation problems.

In hopes of resolving these issues, the tools of loop quantum gravity have been applied to
cosmology in a similar way.  Loop quantum cosmology (LQC) involves the same reduction of the
phase
space as other mini-superspace models; however, the system is quantized using the same techniques as
the full theory of loop quantum gravity.  In other mini-superspace models, the quantization is not
necessarily related to that of a quantum theory of gravity.  This is thought to be responsible for many of
the difficulties that arise.

The results claimed by loop quantum cosmology to date have been impressive.  First of all, it has been
argued that the cosmological singularity is naturally removed in this framework\cite{Sing}.  Secondly,
the initial conditions can be imposed from the behaviour of the equation itself\cite{ICs}.  Finally,
classically divergent quantities are bounded in this framework(e.g. \cite{MLQC}).  This introduces
modifications to the evolution of the universe that seem to produce inflation\cite{Infl} and eliminate
chaos in the Bianchi IX model\cite{B9}.

However, much of the successes of LCQ have come from the semi-classical analysis of the results. 
The interpretation of the quantum states themselves is not entirely clear.  In particular, as with the
Wheeler-DeWitt equation, 'time' does not appear as an explicit variable in the quantum framework.  In
the flat model, one typically picks the scale factor to represent a 'time' variable.  From this, one can
consider the evolution of the other degrees of freedom with respect to that time.  However, in closed
models where the universe recollapses classically, this interpretation seems less intuitive.  In particular,
the same volume corresponds, in general, to two different times.  Furthermore, the classical solution
attains a maximum volume, which one would expect to apply in the quantum case as well.  Therefore,
the probability should decay for large volumes.

Much of the work to date in LQC has involved flat models.  This is not unreasonable since
measurements indicate that we will in a very nearly flat universe.  However, if one wants to address
current cosmological models, one must be capable of addressing the 'flatness problem' associated with
big bang cosmologies.  As such, one must be able to understand the flat model as a limit of the closed
model.

Because of the many issues involving closed cosmological models, it seems necessary to understand the
nature of these solutions in loop quantum cosmology.  In this paper, we will examine the solutions of the
closed model in loop quantum cosmology in the presence of a scalar field.  In
particular, we will focus on the asymptotic behaviour of the model and how one imposes initial
conditions in the exact solutions.  Our analysis suggests that one cannot impose initial conditions that
produce bounded solutions without introducing additional problems.

\section{The Closed Isotropic Model}

Loop quantum gravity is constructed by quantizing the phase space generated by the three-space
connection variable $A$ and its conjugate $E$.  In the isotropic model, these variables can be written
in terms
of the symmetry generators of the space and real multipliers.  The end result is a phase
space defined by the real variables $c$ and $p$ \cite{MLQC} such that
\beq
\{ c,p \} = {1\over 3} \kappa \gamma.
\eeq
where $\kappa = 8\pi G$ and $\gamma$ is the Barbero-Immirzi parameter.  $c$ represents the
degrees of freedom associated with the connection $A$
and is given by,
\beq
\label{c}
c=\Gamma-\gamma{K},
\eeq
where $\Gamma$ is the spin connection and $K$ is the extrinsic curvature.  The spin connection is
related to the intrinsic curvature of the spacial hypersurface, such that $\Gamma=0, {1 \over 2}$ in the
flat and closed models respectively.  The conjugate momenta, $p$, is related to the scale factor $a$
through $|p| = a^{2}$.  With the Hamiltonian constraint,
\beq
H = -{6 \over {\gamma^{2} \kappa}} {sgn(p) \sqrt{|p|}} (c^{2}-c + {{1+\gamma^{2}} \over {4}})
=0
\eeq
this classical phase space describes isotropic cosmologies in general relativity.

When quantizing the phase space of the full theory, one uses smeared variables rather than the basic
variables themselves \cite{QSD}.  Since one would like to follow the full theory as much as possible,
the holonomies of the reduced phase space can be defined in a similar way.  In this case, the integration
can be taken over straight lines because of the symmetries, giving the holonomies of $A$ \cite{MLQC}
as
\beq
\label{holo}
h_{e} = P \exp( \int_{e} A) = \cos(c \mu) + 2 \sin(c \mu) ( {\dot e}^{a} {\omega_{a}^{i}})\tau^{i}
\eeq
where $\mu \in (-\infty,\infty)$ gives the length of the integration (which is not a physical length) and $P$
denotes path ordering.  This leads to an orthonormal
basis of states defined by
\beq
|\mu \rangle = \exp({i \mu c \over 2}).
\eeq
From here, one defines $\hat{p} = -i {l_{p}^{2} \gamma  \over 3} {d \over dc}$ which is an
eigenfunction of these states with eigenvalue $p_{\mu} \equiv {{\mu \gamma l_{p}^{2}} \over 6}$.
From the classical relation $V = |p|^{3/2}$, one can define
\beq
\label{Vol} 
\hat{V} |\mu \rangle = ({|\mu| \gamma l_{p}^{2} \over 6})^{3/2} |\mu \rangle \equiv 
V_{\mu} |\mu \rangle.
\eeq
This is a well-defined operator since $\hat{p}$ is diagonal.  In fact, this can be exploited to produce an
operator for any positive power of $p$.  From these simple operators and the holonomies, one can
produce operator representations of more complicated quantities.  The action of holonomies on the
basis states can be determined by
expanding in imaginary exponentials and using the relation
\beq
\label{exp}
e^{i {c \over 2}} |\mu \rangle = |\mu + 1 \rangle.
\eeq 
Using the above relations, one can define an inverse volume operator
\beq
\label{ivol}
\widehat{a^{-3}}|\mu \rangle = ( {9 \over {j(j+1)(2j+1)} \gamma l_{p}^{2} l}
\sum_{k=-j}^{j}{k 
{p_{\mu+2k}^{l}}} )^{3 \over {2-2l}}|\mu \rangle \equiv d_{j,l}(\mu) |\mu \rangle,
\eeq
where $j$ and $l$ parameterize the ambiguities involved in defining the $V^{-1}$ operator from
classical expressions \cite{amb1,amb2}, where $l \in (0,1)$ and $j$ is a positive integer.   The integer
$j$ arises from our freedom to use different representations of SU(2).  The parameter $l$ arises from
the fact that classically $V^{-1} = (V^{l-1})^{1 \over {1-l}}$.  However, the operators for
$V^{l-1}$ will not preserve this identity on all scales.  Asymptotically, this expression will behave like
$V_{\mu}^{-1}$, for all allowed values of $l$ and $j$, since
\beq
p_{\mu +m }^{l} - p_{\mu - m}^{l} = ({{\mu \gamma l_{p}^{2}} \over 6})^{l}[(1+{m \over
{\mu}})^{l}-(1-{m \over {\mu}})^{l}] \approx ({{\gamma l_{p}^{2}} \over 6})^{l} (2 l m)
\mu^{l-1}
\eeq
when $\mu \gg m$.  However, on very small scales, the inverse volume operator is bounded where the
classical value is not.  The precise behaviour of the eigenvalues on these small scales depends critically
on $j$ and $l$.  

In order to determine the physical states and observables we need to define a Hamiltonian
constraint operator.  One can try to construct the operator along the lines of the full theory
\cite{QSD} where the Euclidian and Lorentzian constraints are quantized separately.  This procedure
was performed in \cite{ILQC} for both the flat and closed models.  For the flat model, this approach
works well; however, in the closed
model, the quantizations is unstable \cite{Stable}.  Therefore another approach was found to be 
necessary for the closed model.

There is a different approach is based on the techniques used for the Bianchi IX model
\cite{HLQC1,HLQC2}.  Since $\Gamma={1 \over 2}$ for the closed model, one can produce an
operator for the Hamiltonian constraint using $c-\Gamma$ in place of $c$.  This method is likely to
have less in common with the full theory since it exploits the form of $\Gamma$, but we can use the
same techniques as the flat case.  The resulting Hamiltonian constraint operator was found to be
\cite{LQCBPI}
\beq
\label{const}
\hat{C}_{grav} = 96 i (\gamma^{3} \mu_{o}^{3} l_{p}^{2})^{-1} ({\sin^{2}(\mu_{o} 
{(c - \Gamma) \over 2}) \cos^{2}(\mu_{o} {(c - \Gamma) \over 2})} -
{\gamma^{2}\mu_o^2 \over 4}{(\Gamma^2 - \Gamma)}) ( \sin(\mu_{o} {c \over 2}) \hat{V} 
\cos(\mu_{o} {c \over 2}) - \cos(\mu_{o} {c \over 2}) \hat{V} \sin(\mu_{o} {c \over 2}) )
\eeq
where the sines and cosines have the same action on states as previously described.  $\mu_{o} \in
(0,\infty)$ is an ambiguity parameter related to the 'length' of holonomies used to construct the
constraint.  The $\Gamma$ terms in the trigonometric operators will be pulled out to form a complex
exponential 
multiplier in the eigenvalue.  As in the full theory, any matter Hamiltonian can be added directed to this
operator to form a full Hamiltonian constraint.

From the constraint operator proposed in (\ref{const}), one can define the physical states of the
system; ie, those that are annihilated 
by the constraint operator.  The obvious starting point is the volume eigenfunctions.  The 
action of the constraint operator on a volume state is
\beq
\label{conmu}
\hat{C}_{grav} |\mu \rangle = 3 (\gamma^{3} \kappa l_{p}^{2})^{-1} {({V_{\mu +\mu_{o}}} -
{V_{\mu - 
\mu_{o}}}) (e^{2 i \Gamma \mu_{o}} |\mu + 4 \mu_{o} \rangle} - {(2 - {4\gamma^{2}\mu_o^2}( 
\Gamma^{2} - \Gamma)) |\mu \rangle} + e^{- 2 i \Gamma \mu_{o}} |\mu - {4 \mu_{o}} \rangle ).
\eeq
Clearly the volume eigenstates, other than the zero volume state, are not annihilated.  
Therefore, we will write any physical state as a sum over volume states via
\beq
\label{state}
| \Psi \rangle  = \sum_{\mu} s_{\mu}(\phi) |\mu \rangle,
\eeq
where $\phi$ denotes the matter dependence.  Therefore, imposing $\hat{C} | \Psi \rangle 
= 0$ yields the condition
\beq
{(V_{\mu +5 \mu_{o}} - V_{\mu + 3 \mu_{o}}) {e^{-i 2 \Gamma \mu_{o}}} s_{\mu + 4 
\mu_{o}}} - {(2 - 4 \mu_o^2\gamma^{2} (\Gamma^2 - \Gamma)) (V_{\mu + \mu_{o}} - V_{\mu 
- \mu_{o}}) s_{\mu }} + (V_{\mu -3 \mu_{o}} - V_{\mu - 5 \mu_{o}}){e^{i 2 
\Gamma \mu_{o}}} s_{\mu - 4 \mu_{o}}\nonumber
\\ = -{1 \over 3} \mu_{o}^{3} \gamma^{3} \kappa 
l_{p}^{2} \hat{H}_{matt} s_{\mu},
\label{conseq}
\eeq
where the matter Hamiltonian, $H_{matt}$, has been added directly to the constraint, as in the full
theory \cite{QSD5}.  This is the discrete equation that describes the state of the universe.  Therefore,
given a matter Hamiltonian, we use (\ref{conseq}) to determine the coefficients of the physical state. 
However, one
should keep in mind that the universe is not in one volume state that 'evolves' according to this 
equation.  The states are only those that include the sum of all volume states.  In the flat model, the
parameter $\mu$ is frequently used to represent 'time'.  However, it is not clear that this interpretation
applies in general.

\section{Exact Solutions}
In order to find exact solutions, we will need to choose a form of matter.  For simplicity, 
we will use a scalar field with the Hamiltonian
\beq
\label{Hmatt}
\hat{H}_{matt} = {1 \over 2} \hat{a}^{-3} p_{\phi}^{2} = -{\hbar^{2} \over 2} 
\hat{a^{-3}} {d^{2} \over {d \phi^{2}}},
\eeq
where $\hat{a}^{-3}$ is defined as in (\ref{ivol}).  We assume that the coefficients have the 
form
\beq
\label{rescls}
s_{\mu} (\phi) = c_{\mu} e^{i \tilde{\omega} \phi},
\eeq
where $\tilde{\omega}$ is a constant and $c$ is independent of $\phi$.  This particular form 
of the matter Hamiltonian and the intial wave function were chosen to make it possible to find a 
solution.  Nevertheless, we can write any possible matter dependence for this Hamiltonian as sum
of solutions of the form (\ref{rescls}).  Using this form, we can 
rewrite the RHS of (\ref{conseq}) as
\beq
\label{constwmatter}
-{1 \over 3} \mu_{o}^{3} \gamma^{3} \kappa l_{p}^{2} \hat{H}_{matt} s_{\mu} = - 
\mu_{o}^{3}\gamma^{3} \l_{p}^{3} \omega^{2} d_{j,l}(\mu) s_{\mu},
\eeq
where
\beq
\label{resclw}
\omega^{2} = {{\tilde{\omega}^{2} \kappa \hbar^{2}} \over {6 l_{p}^{4}}}.
\eeq

We also introduced an exponential containing $\Gamma$ to eliminate the exponentials 
from the Hamiltonian constraint equation via the transformation $s \rightarrow s e^{i \mu 
{\Gamma \over 2}}$.  The resulting equation can be written in iterative form
\beq
\label{inter}
s_{\mu+4 \mu_{o}} = {{l_{p}^3} \over {V_{\mu+5\mu_{o}} - V_{\mu+3\mu_{o}}}} [ (2 
- \gamma^{2} 4 \mu_o^2 (\Gamma^{2} - \Gamma) ) ({{V_{\mu+\mu_{o}}-V_{\mu-\mu_{o}}} 
\over {l_{p}^{3}}})s_{\mu} - \gamma^{3} \mu_{o}^{3} sgn(\mu) 
\omega^{2} l_{p}^{3} d_{j,l}(\mu) s_{\mu}\nonumber
\\ - ({{ V_{\mu-3\mu_{o}} - V_{\mu-5\mu_{o}}} \over {l_{p}^{3}}}) s_{\mu-4\mu_{o}}]
\eeq
which holds for all $\mu \neq -4 \mu_{o}$.  None of these equations depend on $s_{0}$ which drops
out of the equation entirely.  This is why we have no equation for $\mu =-4
\mu_{o}$.  Nevertheless, one can evolve though the classical singularity, $\mu =0$, since choosing
$\mu=0$ in (\ref{inter}) is simply
$s_{4\mu_{o}} = s_{-4\mu_{o}}$.  In solutions where $\mu \neq 4\mu_{o}$, (\ref{inter}) is
always valid; thus, these are well defined around the
zero volume state.  The $sgn(\mu)$ term is introduced because 
one has the freedom to choose the sign on the Hamiltonian constraint.  The $sgn(\mu)$ 
was chosen since there is no reason to expect different solutions in the classically 
disjoint regions $a<0$ and $a>0$.

\begin{figure}
\begin{center}
\includegraphics[width=0.7\textwidth]{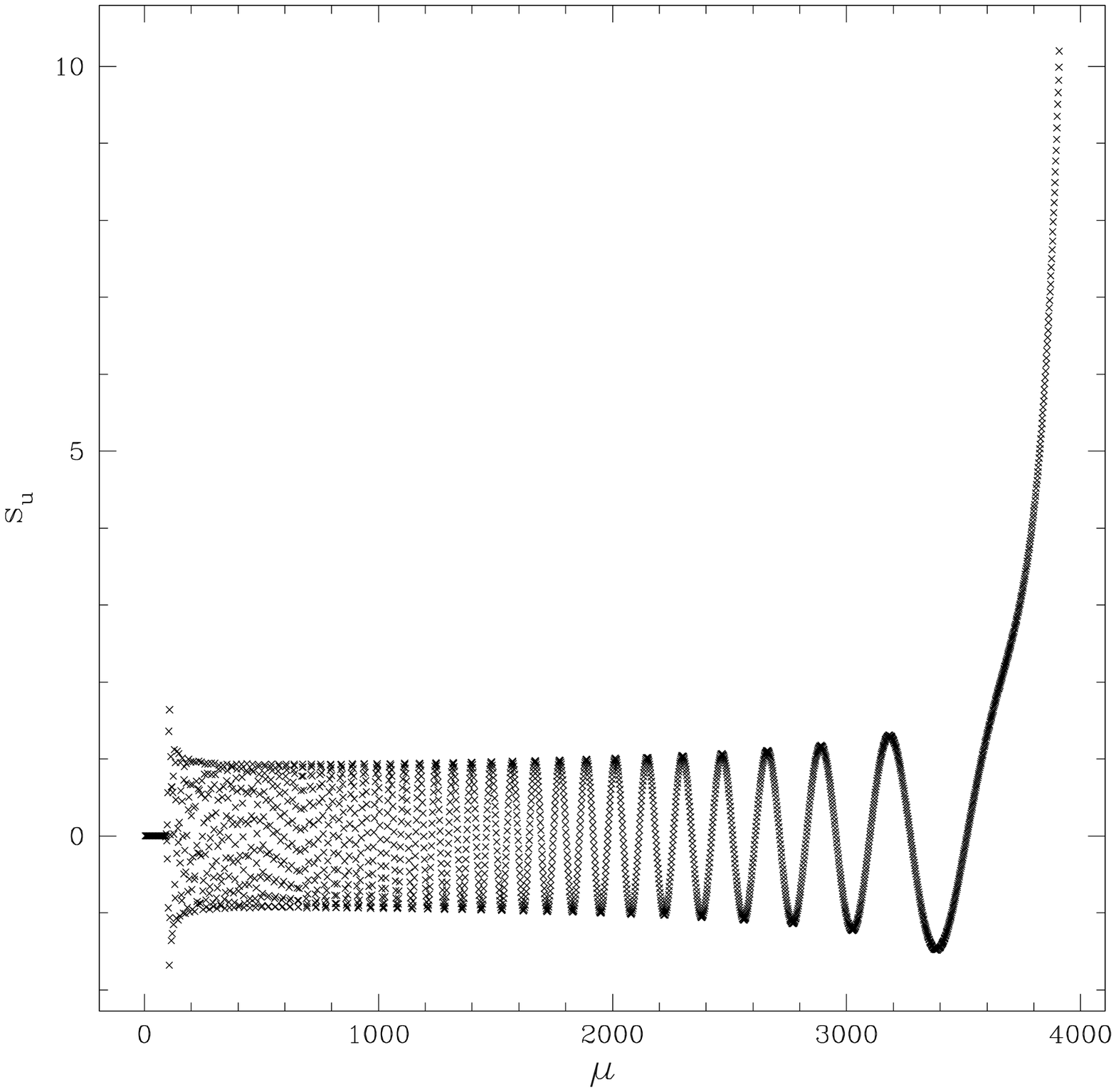}
\caption{This is the exact solution to the closed
$\Gamma ={1 \over 2}$ with $\omega=55$ and $j=20$.
The initial value was set a $\mu=4 \mu_{o}$
with $\mu_{o}=\sqrt{3}/4$.  The state clearly diverges as we
 go to large scales.  Since large $\mu$ is classically disallowed, we would like not to admit these
 states as physical possibilities.  The factor $\exp(i\mu \Gamma /2)$ has been left out of the
solution.}
\label{fig:symm}
\end{center}
\end{figure}

One can solve this equation iteratively by choosing 'initial' conditions at any particular 
point.  In this model, we have 2 degrees of freedom, as long as the solution satisfies the equation
at $\mu=4\mu_{o}$.  However, the form of the solutions shows the first problem with this approach. 
We have factored out the $e^{i \mu {\Gamma \over 2}}$ which will be present in all solutions.  This
factor implies that the state changes rapidly on scales near the plank scale.  These oscillations on small
scales also appear in the flat model and remain to be understood.  However, in the flat model, one
avoids such solutions on the grounds that they do not produce semi-classical physics.  In this case,
since the terms appear explicitly in the equation, we can still approximate (\ref{conseq}) with a
differential equation on large scales.  Therefore, this is only a problem in the sense that it is not clear
what this phase represents and how it relates to quantities that can be 'measured'.

Before we get to the properties of the exact solutions, we will look at the approximate behaviour
on large scales.  This will give a much better qualitative understanding of the solution than 
can be seen from the difference equation itself.

First of all, let us define the new variable $t_{\mu}$ such that
\beq
\label{t}
t_{\mu} = e^{i \mu {\Gamma \over 2}} (V_{\mu +\mu_{o}} - V_{\mu - \mu_{o}}) 
s_{\mu}.
\eeq
This satisfies the simpler equation
\beq
\label{discrete}
t_{\mu + 4 \mu_{o}} - 2 t_{\mu} + t_{\mu -4 \mu_{o}} = - \omega^{2} \mu_{o}^{3} 
l_{p}^{6} d_{j,l}(\mu)  {t_{\mu} \over {({V_{\mu +\mu_{o}}} - {V_{\mu - \mu_{o}}}) }} 
- \gamma^{2}4 (\Gamma^{2} - \Gamma)\mu_o^2 t_{\mu}.
\eeq
When $\mu$ is large, the LHS of (\ref{discrete}) can be approximated by a differential operator,
\beq
\label{diff}
t_{\mu + 4 \mu_{o}} - 2 t_{\mu} + t_{\mu -4 \mu_{o}} = 16 \mu_{o}^{2} {{d^{2}t} 
\over {d \mu^{2}}} = {4 \over 9} (l_{p}^{2} \gamma)^{2} \mu_{o}^{2}{{d^{2}t} \over 
{dp^{2}}}.
\eeq
In the same limit, we have $ d_{j,l}(\mu) = p^{-3/2}$.  The final approximation for the 
difference in the volume eigenvalues.  We can rewrite these in terms of $p$ using,
\beq
\label{volasym}
V_{\mu +\mu_{o}} - V_{\mu - \mu_{o}} = ({|\mu| \gamma l_{p}^{2} \over 6})^{3/2} 
((1+ {\mu_{o} \over \mu})^{3/2} -  (1-{\mu_{o} \over \mu})^{3/2}) \approx ({\gamma 
l_{p}^{2} \over 6})^{3/2} 3 \sqrt{\mu} \mu_{o}
\eeq
and $\sqrt{\mu}= \sqrt{6p \over {\gamma l_{p}^{2}}}$.  From here we get 
\beq
\label{volp}
V_{\mu +\mu_{o}} - V_{\mu - \mu_{o}} = {{\sqrt{p}\gamma l_{p}^{2}} \over 2}.
\eeq 
The end result is the Wheeler-DeWitt-like equation,
\beq
\label{ODE}
p^{2} {2 \over {9 \mu_{o}}} {{d^{2}t} \over {dp^{2}}} + p^{2}2 ({\Gamma^{2} - 
\Gamma}) {t \over {\mu_{o} l_{p}^{4}}} + \omega^{2} t  = 0.
\eeq

In the flat case ($\Gamma = 0$), one can find a solution by assuming $t =A p^{r}$.  
The equation gives the solution
\beq
r = {1 \over 2} \pm \sqrt{ {1 \over 4} - {{9 \omega^{2}} \over 2}}.
\eeq
From this solution, one sees that there is a critical value of $\omega$, $\omega_{c}^{2} = {1
\over 18}$ above which we get oscillatory solutions.  Assuming $\omega > \omega_{c}$, the
solution takes the form
\beq
t=\sqrt{p} \lgroup B\exp(+iR \log(p))+C\exp(-iR \log(p)) \rgroup ,
\eeq
where $R$ is the imaginary part of $r$.  Recall that we absorbed volume terms into $t$.  From
(\ref{volp}), we find that $s \propto {t \over {\sqrt{p}}}$.  Thus, in the large-scale limit, $s$ is purely
oscillatory.  Similar analysis has been performed for the flat model using quantizations of the Euclidian
\cite{ILQCM} and the Lorentzian \cite{ILQCM2} constraints and yield the same critical value.

In the closed case ($\Gamma = {1 \over 2}$), one can drop the matter term for large $p$ since it
drops off as ${1 \over {p^{2}}}$ relative to the other terms.  The resulting equation has simple
solutions in terms of exponentials,
\beq
\label{asysol}
t(\mu) = B \exp(-{1 \over 4} \gamma \mu) + C \exp(+{1 \over 4} \gamma \mu)
\eeq
where $B$ and $C$ are constants.  The real exponentials are not surprising, given that 
large values of $a$ are classically disallowed.  Like any conventional quantum mechanics 
problem, one would like to impose the boundary condition that $t \rightarrow 0$ as $|\mu| 
\rightarrow \infty$.  Clearly one would set $C = 0$ for large positive $\mu$, and $B=0$ for 
negative $\mu$.  However, the iterative equation defines solutions for both positive and negative $\mu$.

Let us consider a solution where we pick some initial value $s_{4 \mu_{o}} = D e^{i \tilde{\omega}
\phi}$.  From 
(\ref{inter}), we find that
\beq
\label{zero}
s_{ 8 \mu_{o}} = {{l_{p}^3} \over {V_{9 \mu_{o}} - V_{7 \mu_{o}}}} [ (2 + 
\gamma^{2}\mu_o^2) ({{V_{5 \mu_{o}}- V_{3 \mu_{o}}} \over {l_{p}^{3}}}) -\gamma^{3} 
\mu_{o}^{3} sgn(\mu) \omega^{3}l_{p}^{3} d_{j,l}(\mu)] s_{4 \mu_{o}}.
\eeq
Since $s_{8 \mu_{o}} \propto D$, $D$ will scale all values of the solution for positive $\mu$. 
Furthermore, when we set $\mu=0$, the constraint equation 
turns into $s_{4 \mu_{o}} = s_{-4 \mu_{o}}$.  The evolution is symmetric in $\mu$ so 
$D$ will be present for all negative $\mu$ as well.  Therefore, the solution will not 
depend on how we choose our initial value.  One such exact solution is shown in figure \ref{fig:symm}.  
As one can see, the solution diverges at large $\mu$.  Since we can only set an overall 
normalization, the only solution we have (other than $|\Psi \rangle = 0$) is unphysical.  

\begin{figure}
\begin{center}
\includegraphics[width=0.7\textwidth]{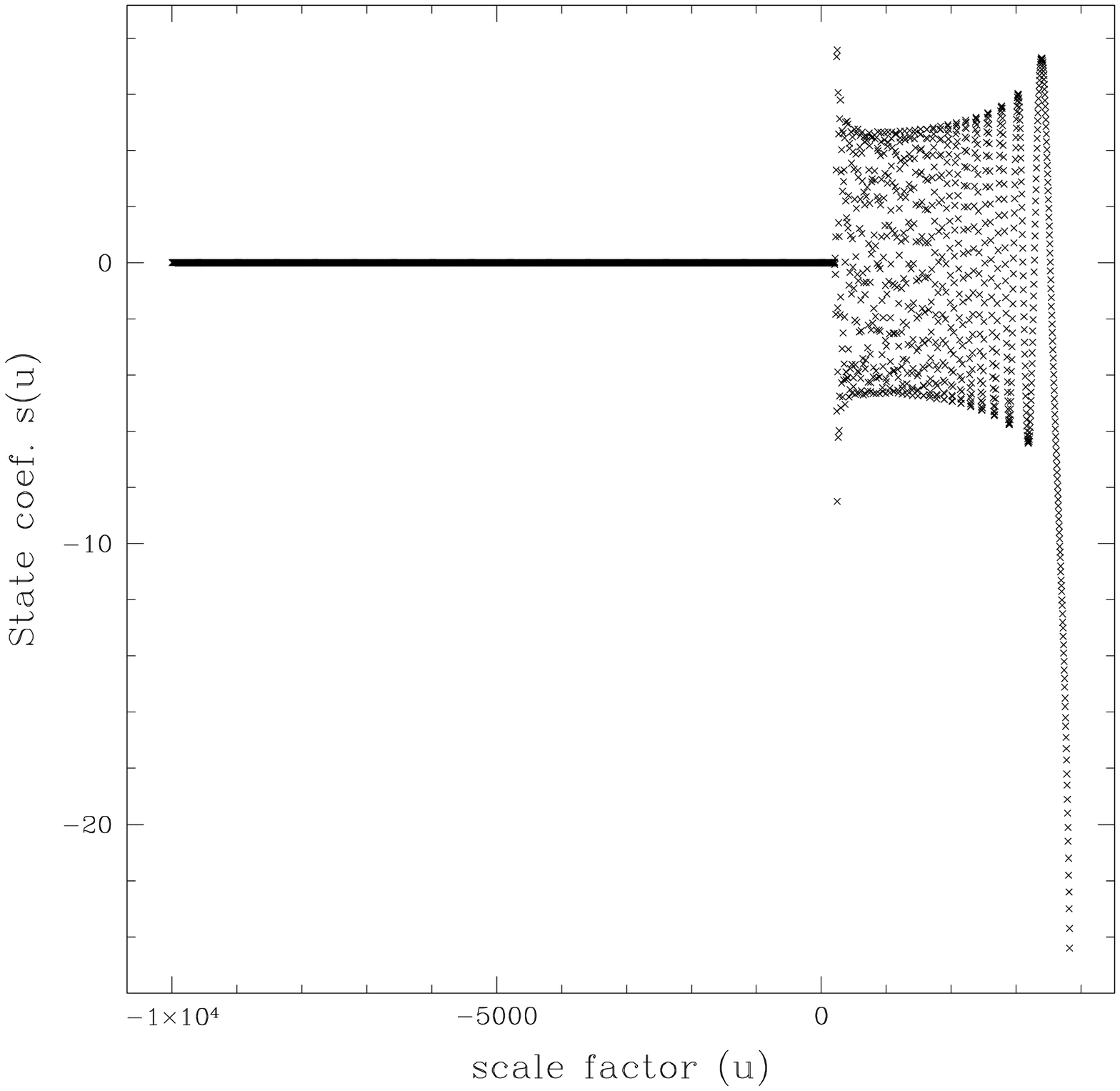}
\caption{This is the exact solution to the closed 
model $\Gamma =1/2$ with $\omega=55$ and $j=20$.  The 
initial value was set a $\mu= 3 \mu_{o}$ with $\mu_{o}=1$.  
The initial conditions have chosen to eliminate the growing solution in 
the $-\mu$ domain.  However, it is clear we cannot eliminate the growing 
solutions in both positive and negative domains for the asymmetric model 
(i.e. the model where we don't hit zero volume).  The factor $\exp(i\mu \Gamma /2)$ has been
left out of the solution.}
\label{fig:asymm}
\end{center}
\end{figure}

In the case where we define our initial value at $\mu \neq n4\mu_{o}$, where $n$ is an integer, there is
no additional constraint on the solution from the zero volume state.  However, because we no longer
have the zero volume to eliminate terms, we do not have symmetric evolution in positive and negative
$\mu$.  Therefore, if we eliminate the divergence in the negative domain, we will still have the 
problem in the positive domain and vice versa.  Figure \ref{fig:asymm} shows the state of the universe
when we impose that $s_{\mu} \rightarrow 0$ for negative $\mu$.  Clearly, the problem remains.  This
is
very similar to the problem of producing the pre-classical solution in the flat model with the
Lorentzian constraint \cite{ILQCM2}.  In that case, one can only produce smooth solutions in either
$\mu < 0$ or $\mu >0$, but not both simultaneously.

Solutions to this particular model have been presented \cite{LQCBPI} for the case where a
cosmological constant is included.  The solution avoids the difficulties presented here.  In particular, one
can rewrite the matter Hamiltonian as
\beq
\label{Cosconst}
H_{matt}=a^{-3} p_{\phi}^{2} + a^{3} V_{o}
\eeq
where $V_{o}$ is the cosmological constant.  The divergent solutions described above are not
observed in this case because exponentially growing solutions are no longer a feature of the asymptotic
solution.  In particular, this introduces an additional term which dominates at large $\mu$.  The resulting
behaviour is described by
\beq
\label{ODECC}
{d^{2}t \over {d \mu^{2}}} + |\mu| V t =0
\eeq
where $V = V_{o} {1 \over 48} \mu_{o} \gamma^{3} \kappa l_{p}^{2}$.  The solutions to
(\ref{ODECC}) are defined in terms of Airy functions
\beq
t=A'Ai(-V^{1/3}|\mu|)+B'Bi(-V^{1/3}|\mu|),  
\eeq
where $A',B'$ are constants.  For $V_{o}>0$, the solutions are well behaved for all $\mu$. 
Coincidently (or perhaps not), this is the same condition that determines if large $a$ is classically
disallowed.  For a generic matter Hamiltonian and wave-function $\phi$ and $a$ will not decouple, but
there will be regions of the $a-\phi$ plane that are classically disallowed, where we expect similar
problems will arise.

There is reason to be concerned that the numerical results presented in figures \ref{fig:symm} and
\ref{fig:asymm} do not represent the exact solutions of the constraint equation.  Because the solutions
will take the form of exponentially growing and decaying solutions, numerical round-off errors may
grow exponentially.  This problem was tested explicitly.  First of all, the solutions calculated using
between 16 and 1000 digits all behaved in a similar manner and were within a few percent of each
other where the divergences become large.  Secondly, one can produce solutions that are decaying
where solutions usually diverge by avoiding the constraint at zero (or leaving the opposite $\mu$
domain to diverge).  These solutions are qualitatively different and near $\mu=0$ the ratio $s_{8
\mu_{o}} / s_{4 \mu_{o}}$ differ by orders of magnitude.  This behaviour has been tested over a
reasonable range in $\gamma$, $\mu_{o}$, $l$, $j$ and $\omega$ with similar behaviour in all cases.  

\begin{figure}
\begin{center}
\includegraphics[width=0.7\textwidth]{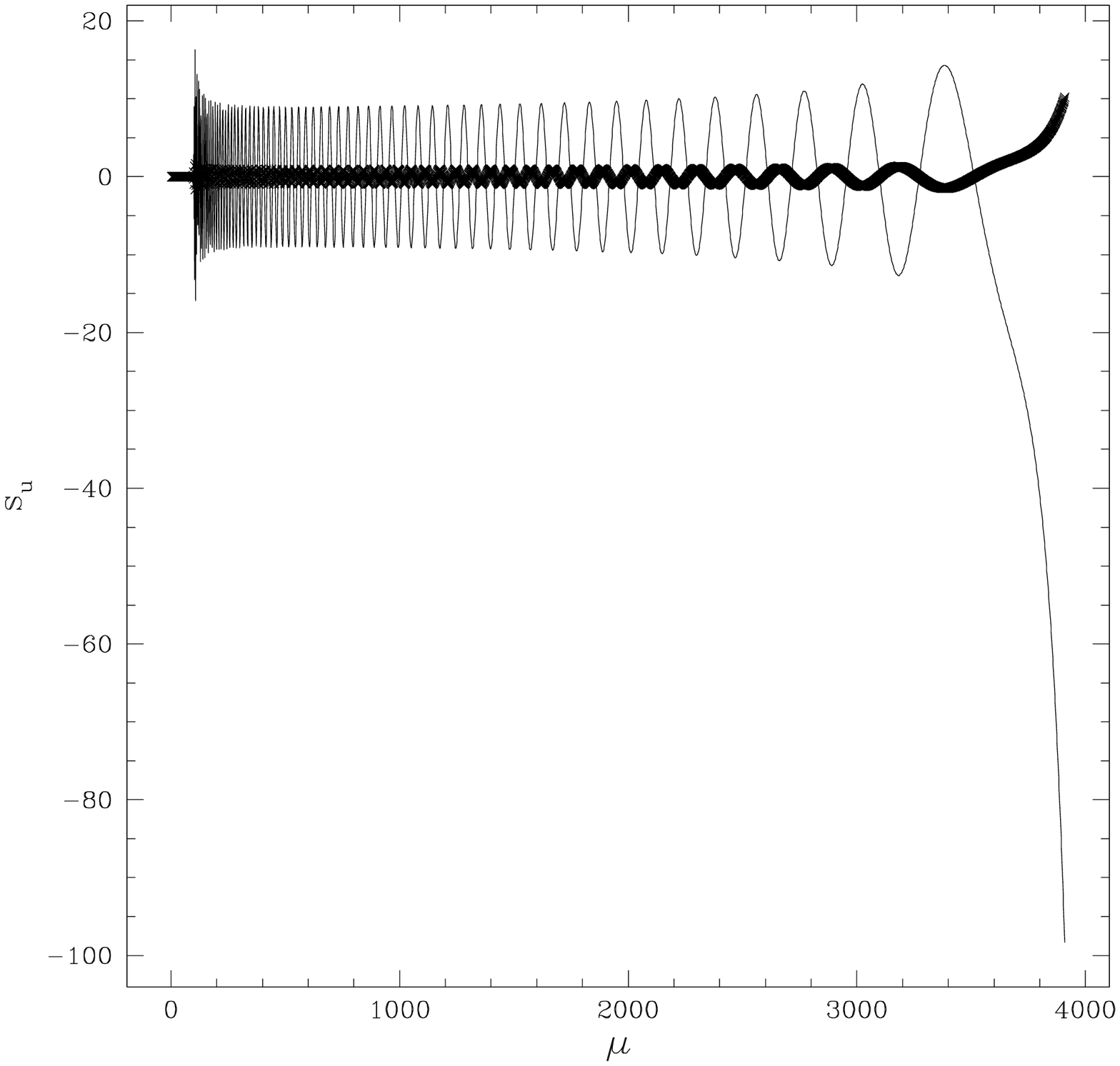}
\caption{This is the exact solution to the closed 
model $\Gamma =1/2$ with $\omega=55$, $j=20$ and $\gamma={{\log{2}}\over {\sqrt{3} \pi}}$. 
The crosses represent the solution for $l=.45$ and solid line is for $l=.4475$.  The initial value is set at
$\mu=4\mu_o$ and is equal in both cases.  The change in the sign of the growing exponential part of
the solution suggests that we can fix $l$ such that only the decaying solution remains.}
\label{fig:w55}
\end{center}
\end{figure}

Although these ambiguity parameters provide additional freedom in our
solutions, it is unclear whether they can cure our problems or whether
they simply relocate them.  First of all, there have been arguments
presented that should fix our choice of both $\gamma$ and $\mu_{o}$
\cite{MLQC}.  Assuming these to be fixed, the only remaining
parameters are $j$ and $l$.  If we fix all values and adjust $l$ over
the allowed range $(0,1)$, there are changes in the sign of the
diverging exponential for fixed $s_{4\mu_{o}}$.  This suggests that we
can fine tune $l$ such that the exponentially growing solution
vanishes.  Unfortunately, this fine-tuned value depends on $\omega$.

\begin{figure}
\begin{center}
\includegraphics[width=0.7\textwidth]{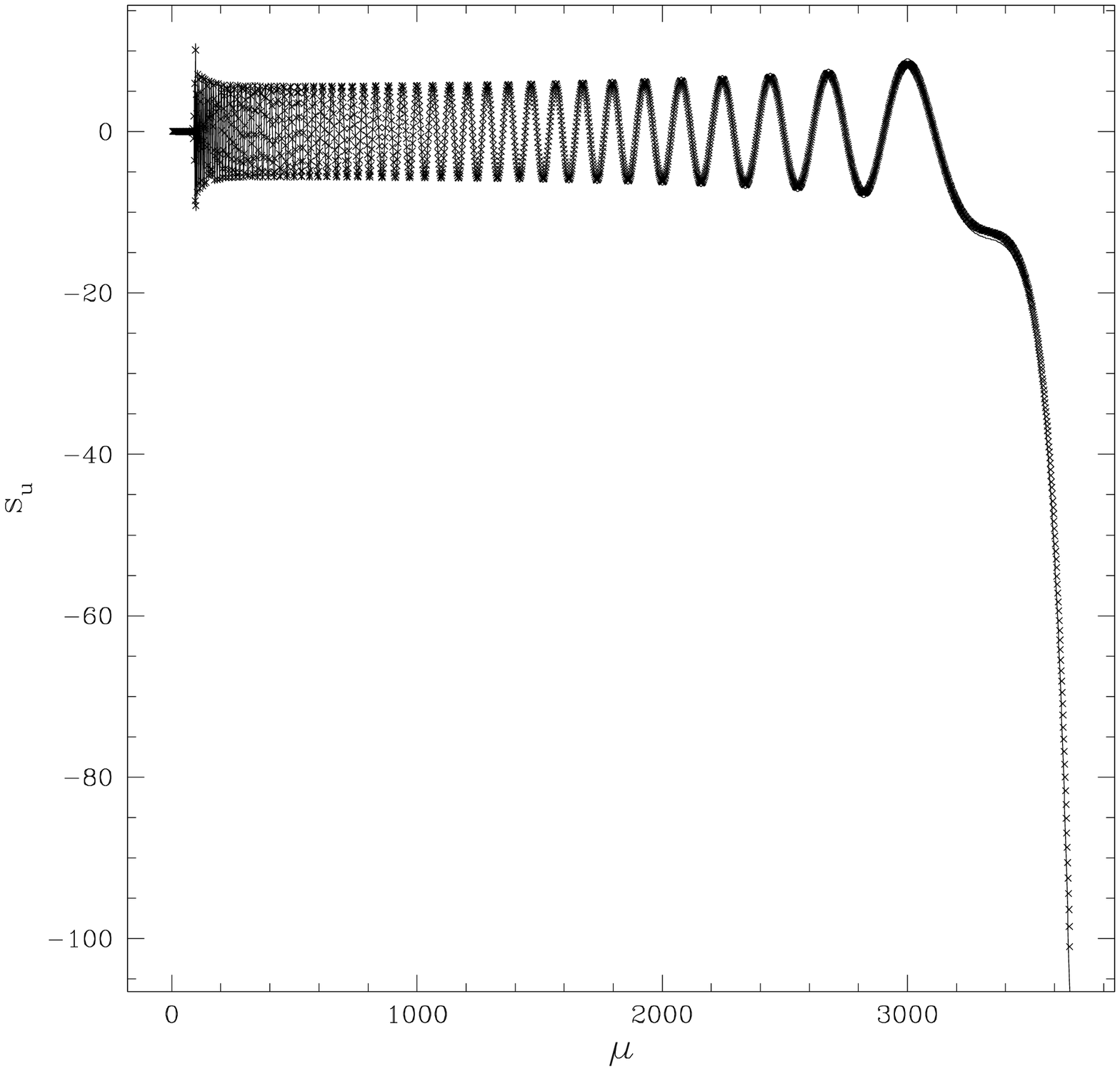}
\caption{This is the exact solution to the closed 
model $\Gamma =1/2$ with $\omega=50$, $j=20$ and $\gamma={{\log{2}}\over {\sqrt{3} \pi}}$. 
The crosses represent the solution for $l=.45$ and solid line is for $l=.4475$.  The initial value is set at
$\mu=4\mu_o$ and is equal in both cases.  These solutions are nearly identical for all $\mu$.  This
suggests that there is no single value of $l$ that will produce bounded solutions for all choices of
$\omega$. }
\label{fig:w50}
\end{center}
\end{figure}

In figure \ref{fig:w55}, we have an example of the sign flip that occurs for a
very small change in $l$ when $\omega=55$.  However, if $\omega=50$,
the solutions for the same two values of $l$ are virtually
indistinguishable, as seen in figure \ref{fig:w50}.  Further choices of $l$ seem to suggest that there is no
common $l$ in that range for both $\omega=50$ and $\omega=55$
that eliminates the unbounded solution.  The same holds for a wide
range of choices of $\omega \leq 500$, $j \leq 1000$ and $\gamma \leq 1$.  Since
the inverse density operator is defined on the kinematic Hilbert
space, its definition should not depend on the form of the matter
Hamiltonian or the particular state of the matter.  In this particular
model, one should be able to expand the solution as a sum over all
$\omega$.  One should not have to define $l(\omega)$, before doing so.
Therefore, it seems unlikely that a choice of these parameters produce
an acceptable model.  

\begin{figure}
\begin{center}
\includegraphics[width=0.7\textwidth]{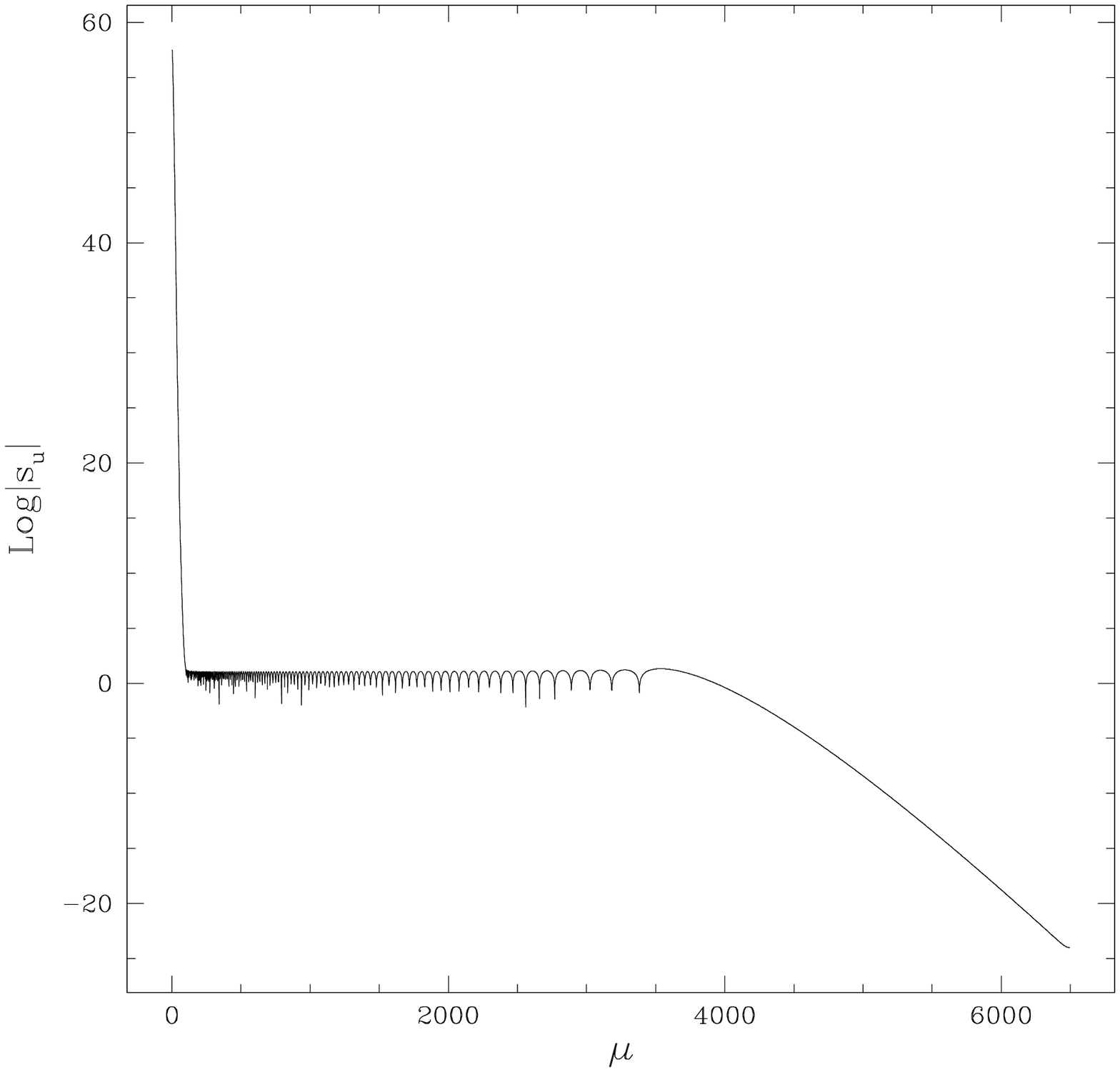}
\caption{This is a plot of the log of modulus of $s_{\mu}$ for the closed 
model $\Gamma =1/2$ with $\omega=55$, $j=20$ and
$\gamma={{\log{2}}\over {\sqrt{3} \pi}}$.  The crosses represent the
critical solution for $l=.447....$ in which the solution is bounded at
large scales.  The state is normalized such that the amplitude of the
oscilation is of the same order as the other solutions.  Thus, the
value at $\mu=4\mu_o$ is nearly $10^{60}$ compared to $10^{-50}$ in the
other solutions shown.  Clearly, the behaviour near $\mu=0$ dominates
the solution.  }
\label{fig:crit}
\end{center}
\end{figure}

The bounded solutions achieved by fine-tuning $l$ have the feature of
being sharply peaked around $\mu=0$, as suggested by figure \ref{fig:crit}.  This is an additional
difficulty
of the fine-tuned solutions.  Not only does one need to fix
$l(\omega)$, but one also needs to explain to meaning of the
solutions.  In particular, one should address why the only states are
dominated by the smallest volume eigenstates.

Beyond fixing $l$, we could also fix other parameters in hopes of
producing bounded solutions.  For example, one could adjust the initial
point (instead of assuming the symmetric solution) to some critical $\mu
\in x4\mu_o$ where $x \in (0,1)$.  This would eliminate the continuous
spectrum of solutions for fixed $\omega$, etc in the same way we fixed
$l$ or any of the ambiguity parameters.  However, any one of these choices
would depend on all the other parameters in the theory.  Therefore, we
can restrict one variable in order to produce bounded solutions but at
the cost of introducing additional matter and ambiguity
dependences into the states and the constraint.  Therefore, one needs to ask whether it
is sensible to have such a theory.

\section{Discussion and Conclusion}

In the analysis of the behaviour of a massless scalar field in the closed model from ILQC, it seems that
divergent behaviour of the solutions at large scales is unavoidable.  In this analysis, we consider the
solutions for a wide range of parameters (e.g. $j$, $l$).  In every case, the solution oscillates at small
scales but grows exponentially for large $\mu$.  Attempts to constrain the ambiguity parameters
through the boundary conditions seems to require significant matter dependence, and thus is an
undesirable solution, if it works at all.

When trying to produce a closed model using techniques from the full theory, one gets an unstable
discrete equation.  If the symmetry is used to make the constraint simpler to quantize, we get
unbounded coefficients for the volume states in classically disallowed regions.  If we are to interpret
these solutions as representing a probability, then clearly we cannot allow the latter quantization.  This
assumption is not completely trivial.  The physical meaning of these states is still unknown.  Without an
inner product and observables, the interpretation of the solution is not entirely clear (one 
possible interpretation can be found in \cite{Inter}).  Nevertheless, it
seems unlikely that one can find an interpretation where the large-scale divergences are appropriate. 

We are left with the problem of how to quantize the closed model in loop quantum gravity.  From this
point, it seems unlikely that a minor modification to the techniques applied to the flat model will be
successful.  Therefore, one needs to reconsider the techniques used from the first step in this procedure. 
One of the problems is that the mini-superspace treatment has removed the coordinate freedom one
usually uses to eliminate the spin connection.  If this is indeed the case, there must be a way to construct
the cosmology from the full theory that does not suffer from this problem.  Hence, the full theory might
avoid these difficulties.  This would also suggest that this mini-superspace model has less in common
with the full theory than was intended.

Perhaps the most interesting aspect of the divergences that occur in this model is that they seem to
occur under the same conditions that lead to a re-collapsing universe.  The re-collapsing universe has
typically been difficult to interpret in quantum cosmology.  In this framework, it would be very difficult
to understand the 'evolution' of a re-collapsing universe in terms of the volume states.  It is interesting
that our quantization seems to fail in the same place our framework would lead to significant
interpretation problems.

It remains to be seen whether these types of problems rise in the Bianchi IX model from which this
isotropic quantization was derived \cite{HLQC2}.  The homogenous solutions contain more degrees of
freedom but will also have additional constraints and boundary conditions.  One might hope that the
problems found here result from limiting the phase space to classically isotropic solutions.  If the
homogenous solutions avoid these divergences then it would suggest that we might not need to worry
about the difficulties in the isotropic case.

\section{Acknowledgements} 

The DG would like to thank Bojan Losic for his comments and
suggestions over the course of this work.  WGU thanks the CIAR for
support and both thank NSERC for their support of this work.


\begin{thebibliography}{99} 
\bibitem {Sing} M. Bojowald, Phys. Rev. Lett. 86, 5227 (2001) 
\bibitem {ICs} M. Bojowald, Phys. Rev. Lett 87, 121301 (2001)
\bibitem {Infl} M. Bojowald, Phys. Rev. Lett 89, 261301 (2002) 
\bibitem {B9} M. Bojowald and G. Date, Phys. Rev. Lett 92, 071302 (2004) 
\bibitem {QSD} T. Thiemann, Class.Quant.Grav. 15, 839 (1998)
\bibitem {QSD5} T. Thiemann, Class.Quant.Grav. 15, 1281 (1998)
\bibitem {MLQC} A. Ashtekar, M. Bojowald and J. Lewandowski, Adv.Theor.Math.Phys. 7, 233
(2003)
\bibitem {LQCBPI} M. Bojowald and K. Vandersloot, Phys. Rev. D67, 124023 (2003)
\bibitem {ILQC} M. Bojowald, Class.Quant.Grav. 19, 2717 (2002)  
\bibitem {Stable} M. Bojowald and G. Date, Class.Quant.Grav. 21, 121 (2004)
\bibitem {amb1} M. Bojowald, gr-qc/0402053
\bibitem {amb2} M. Bojowald, Class.Quant.Grav. 19, 5113 (2002) 
\bibitem {ILQCM2} F. Hinterleitner, and S. Major, Phys. Rev. D 68, 124023 (2003)
\bibitem {HLQC1} M. Bojowald, Class.Quant.Grav. 20, 2595 (2003) 
\bibitem {HLQC2} M. Bojowald, G. Date and K. Vandersloot, Class.Quant.Grav. 21, 1253 (2004)
\bibitem {ILQCM} M. Bojowald and F. Hinterleitner, Phys.Rev. D66, 104003 (2002)
\bibitem {Inter} D Colosi and C Rovelli, Phys.Rev. D68 104008 (2003) 
\end{thebibliography}
\end{document}